\documentclass[12pt]{article}

\usepackage{amsfonts, amsmath}

\newcommand{\be}{\begin{equation}} \newcommand{\ee}{\end{equation}}

\usepackage{latexsym}

\begin{document}

\title{Entropy in the Present and Early Universe}

\author{Alexander E. Shalyt-Margolin}
\date{}
\maketitle
\begin{abstract}This is a short analysis of the changes in the concept of
entropy as applied to physics of the present-day and Early
Universe. Of special interest is a leading role of such a notion
as deformation of a physical theory. The relation to a symmetry of
the corresponding theory is noted. As this work is not a survey,
the relevant author's works are mainly considered. This  paper is
to be published in special issue "Symmetry and Entropy" of journal
\textbf{SYMMETRY: Culture and Science}
\end{abstract}

\section{Introduction}
Two principal theories which are revolutionary for the fundamental physics
-- quantum mechanics and relativity theory - were created in the XX-th
century. The third theory including statistical mechanics and thermodynamics
has been formed as far back as the XIX-th century. By the eighties of the
last century all these theories have acquired their complete form. They were
effectively used, actually enabling solution of all the known problems
within the scope of the existing paradigm. However, energies are constantly
increasing with the development of the physical experiment. At the present
time CERN is testing the operation a Large Hadronic Collider (LHC) that
could significantly widen the opportunities for detection of new particles
and previously unknown phenomena. Besides, in the last decade a new unique
field of theoretical physics - cosmological microphysics - has been
formed just at the junction of elementary particle physics and cosmology
and, due to the advances in astrophysics and modern technology (space
probes, Hubble telescope, etc.), has regained the experimental status. In
the end, all these makes it possible to tackle a number of problems
associated with the early Universe (instants immediately after the Big
Bang).

Therefore, of critical importance is an extension of the available
fundamental physics paradigm to find the answers for the questions
concerning the early Universe as all the above-mentioned theories work well
on the scale of the energies known in the every-day life or in modern
accelerators, failing with the energies of the Big Bang. Specifically, the
general relativity theory in this case seems to lose its force of
prediction: it is unknown what happens with the space-time at Planck's
scales ($10^{-33}cm)$, most likely ceasing to be continuous. A quantum
theory should be modified in a similar way, in particular, due to the
divergence problems preventing the correct results in the region of higher
energies. A new paradigm should extend the old one rather than ruin it. Any
theory in physics has its applicability limits. For all the basic theories
these limits are associated with some energy limits. Because of this, the
paradigm should expand the limits so that in the low-energy limit the
well-known theories could accurately provide the result experimentally
checked over and over.

\textbf{With my paper I would like to pay a tribute to my first
Since Manager Professor Alexandre Zalesski who will be celebrating
his 70-th anniversary in January 2009.}

\section{``Deformed'' physical theories }
As indicated above, any new paradigm is a change in the old one to extend
its applicability limits. This is true not only in physics. In every field
of science or culture, a new paradigm arises when considering the problem
that, as a rule, necessitates widening of the initial scope for the
successful solution. Such a situation is typical for the development of a
new method, concept or idea.

In physics a new paradigm is usually related to new constants and variables.
To illustrate, the ``advent'' of a new constant (Planck's constant $\hbar )$
in quantum mechanics as compared to the classical one was followed by the
development of a completely new mathematical apparatus. It can be said that
the classical mechanics appears from the quantum mechanics because of
passage to the limit $\hbar \to 0$.

There is a similar passage from nonrelativistic to relativistic dynamics
\textbf{Faddeev (1989, p.15)} associated with a change in the movement group
of space-time: from Galilean transformations to those of
Lorentz-Poincar\'{e}. Both these groups are ten-parametric.

Galilean transformation is as follows
\begin{equation}
\label{eq1}
\vec {x}\to \vec {x}+\overrightarrow v t,
\quad
t\to t\quad ,
\end{equation}
and in the case of Lorentz transformations we have
\begin{equation}
\label{eq2}
\vec {x}\to \frac{\vec {x}+\overrightarrow v t}{\sqrt {1-v^2/c^2} }
\end{equation}
\begin{equation}
\label{eq3}
t\to \frac{t+\left( {\mathop v\limits^\to \cdot \mathop x\limits^\to }
\right)/c^2}{\sqrt {1-v^2/c^2} }\quad ,
\end{equation}
where $\vec {x}$ are the coordinates, $t$ is the time, $\overrightarrow v $
is the relative velocity of the reference system in motion.

As compared to Galilean transformations, those of Lorentz involve the fixed
constant $c$ - speed of light - that is equal to $3\cdot 10^{10}cm/s$.
When $c\to \infty $, expressions in (2) change over to (\ref{eq1}), i.e. Lorentzian
transformations become those of Galileo (and similarly, the corresponding
groups and their Lie algebras).

Now we introduce the definition of a ``deformed'' physical theory.

\textbf{Definition 1.} Deformation of a physical theory is understood as its
extension due to the introduction of one or several additional parameters so
that the initial theory appears on passage to the limit.

In terms of Definition 1, the relativistic dynamics group represents
deformation of the group for nonrelativistic motion, where the deformation
parameter is $1/c^2$ \textbf{Faddeev (1989, p.15).}

In a similar way, quantum mechanics represents ``deformation'' of the
classical one with the deformation parameter $\hbar $ (or $\frac{1}{\hbar
})$. In the following section this example will be treated in greater
detail. Now, using \textbf{Faddeev's} approach\textbf{, }we consider such a
notion as stable deformation: \textbf{Faddeev (1989, p.15) }``{\ldots} In a
mathematical theory of the algebraic structure deformations there is a
notion of the stable structure. It is said that a structure is stable when
all similar deformations are equivalent to this structure.''. ``From this
viewpoint, quantum mechanics is stable as opposed to the classical mechanics
allowing for the nonequivalent deformation -- quantum mechanics''. Here by
``similar'' we mean the structures with a deformation parameter close in its
value to the initial one \textbf{Gerstenhaber (1964).}

\textbf{Definition 2.} Stable deformation of a physical theory is recognized
as the deformation when all the like (i.e. having similar deformation
parameters) deformations are equivalent.

Despite the fact that in quantum mechanics $\hbar $ is a constant equal to
$\simeq 10^{-27}$gcm$^{2 }$s$^{-1 }$one can easily imagine ``quite a
number'' of quantum mechanics with the deformation parameters close to
$\hbar $ but still nonzero, which are equivalent mathematically.

The third principal achievement of physics in the XX-th century is the
general relativity theory -- stable  deformation in terms of
Definition 2 meaning that \textbf{Faddeev (1989, p.16).} ``a gravitational
theory is just based on the replacement of the flat Minkowski space-time by
the common-position curved pseudo-Riemann space. It can be argued that in a
set of such spaces the flat space is a degeneracy, whereas spaces in the
neighborhood of the curved one are also curvilinear. A measure of
deformation is determined by the gravitation constant $G$, inherited from
Newton and introduced into Hilbert-Einstein equation. The dimensionality of
this constant is functionally independent of $\hbar $ and $c$, and together
with them form the basis for all dimensional parameters.''

\section{``Deformed'' quantum mechanics of the Early Universe. Different
approaches. }
As noted in the previous section, contrary to the classical mechanics,
quantum mechanics involves an additional parameter ( $\hbar )$ in the region
of well-known energies. But at the energies close to those of the Big Bang
this hardly is so. An important role of Planck's quantities in this case is
revealed. The Planck's length is given by
\begin{equation}
\label{eq4}
\ell _p \equiv \sqrt {\frac{G\hbar }{c^3}} \simeq 1.6\cdot
10^{-35}m=1.6\cdot 10^{-33}cm,
\end{equation}
the Planck's time is found from
\begin{equation}
\label{eq5}
t_p \equiv \frac{\ell _p }{c}=\sqrt {\frac{G\hbar }{c^5}} \simeq 0.54\cdot
10^{-43}s,
\end{equation}
the Planck's mass is obtained as
\begin{equation}
\label{eq6}
M_p \equiv \frac{\hbar }{c\ell _p }=\sqrt {\frac{\hbar c}{G}} \simeq
2.2\cdot 10^{-8}kg,
\end{equation}
and the Planck's energy is determined as
\begin{equation}
\label{eq7}
E_p =\sqrt {\frac{\hbar c^5}{G}} \simeq 1.2\cdot 10^{19}GeV.
\end{equation}
Planck's quantities are understood as the known lengths, times and the like
represented by the fundamental constants only. As follows from (\ref{eq4}) -- (\ref{eq7}),
Planck's quantities are defined by three fundamental constants ($\hbar ,$
$c$, and $G)$ rather than one.

In this way a quantum theory (mechanics) of the early Universe should
involve all these fundamental constants $\hbar , \quad c$, $G$ as the principal
energies associated with the processes proceeding in the early Universe are
comparable to the Planck energy$E_p $.

Next we show how the well-known quantum mechanics should be deformed to
involve the above-mentioned three constants.

\subsection{Heisenberg uncertainty relations, generalized uncertainty
principle, and deformation of Heisenberg's algebra.}

>From the classical mechanics it is known that for any particle one can
measure an exact value of its coordinate $x$ and momentum $p$.

But in quantum mechanics this is not the case any more. As demonstrated by
\textbf{Heisenberg (1927), }the accuracy has a natural limit: the greater
accuracy of the coordinate measurements we have the less accurate momentum
measurements we get, and vice versa
\begin{equation}
\label{eq8}
\Delta x\cdot \Delta p\ge \hbar \quad ,
\end{equation}
or equivalently
\[
\Delta x\ge \frac{\hbar }{\Delta p}.
\]
Expression (\ref{eq8}) represents one form of the commonly known uncertainty
relations (or uncertainty principle) of Heisenberg who in the process of
their derivation has assumed that elementary particles take part only in
electromagnetic interactions. This assumption stands to reason since at the
known energies the electromagnetic interactions are higher than the
gravitational ones by several orders.

Note that in literature (\ref{eq8}) is commonly used in the form
\[
\Delta x\cdot \Delta p\ge \frac{1}{2}\hbar .
\]
But in the early Universe, at Planck's scales, gravitational interaction
would be comparable to the electromagnetic ones and should be included.
Because of this, relation (\ref{eq8}) must be modified so that it can occur in the
low-energy limit. According to some modern theories (e.g., string theory)
\textbf{Kaku (1988),} relations (\ref{eq8}) at the Planck's energies $(\sim E_p )$
should be modified as follows \textbf{Veneziano (1986)},
\textbf{Witten (1996)}
\begin{equation}
\label{eq9}
\Delta x\ge \frac{\hbar }{\Delta p}+\gamma \ell _p^2 \frac{\Delta p}{\hbar
},
\end{equation}
where $\gamma $ is a certain dimensionless coefficient. This modification is
referred to as the Generalized Uncertainty Principle (GUP)

Inequality (\ref{eq9}) has been derived beyond the string theory as well
\textbf{Adler, Santiago, (1999) }.

On change-over to low energies, the second term in the right side
of (\ref{eq9}) becomes negligible modifying relation (\ref{eq9})
to (\ref{eq8}). At the same time, one should remember that, in
contrast to the classical mechanics, coordinates and momenta in
quantum mechanics are not numbers but operators, mathematically
represented by matrices (infinite in the general case) generating
the Heisenberg algebra \textbf{Dirac (1958), Messiah (1965) and
given by the commutation relations}
\begin{equation}
\label{eq10}
\left[ {q_i ,q_j } \right]=0,
\quad
\left[ {p_j ,p_k } \right]=0,
\quad
\left[ {q_j ,p_k } \right]=i\hbar \delta _{jk}
\end{equation}
As usual, in formula (\ref{eq10}) $q_i $ is the operator of the $i$-th coordinate,
$p_j -j$-th momentum, $\delta _{jk} $ is the Kronecker delta, and $\left[
{a,b} \right]=ab-ba$.

Consequently, deformation of quantum mechanics in the Early Universe should
be Heisenberg's algebra deformation but including all three fundamental
constants.

All these deformations in the low-energy limit should lead to commutation
relations (\ref{eq10}). Nonequivalent deformations may be numerous in accordance with
various sequences converging to one and the same limit \textbf{Kempf,
Mangano, Mann (1995), Maggiore (1993)}.

To illustrate, two numerical sequences with a common term $a_n =1/n$ and
$b_n =(-1)^n/n^2$ at $n\to \infty $ have the same limit $0$, still being
absolutely different. Also, it should be noted that a common feature of all
the deformations mentioned is the noncommutativity (or nonpermutable nature)
of the operators with coordinates $q_i $ and $q_j $ for different $i$ and$j$
. This means that for the high-energy deformation of Heisenberg algebra we
have
\begin{equation}
\label{eq11}
\left[ {q_i ,q_j } \right]\ne 0.
\end{equation}
Relation (\ref{eq10}), $\left[ {q_i ,q_j } \right]=0$, appears on going to low
energies, therefore in the low-energy limit, where the quantum gravitational
effects are negligible, noncommutativity (\ref{eq11}) is of no importance.

\subsection{Density matrix deformation at Planck's scales }

There is an alternative approach to the deformation of quantum mechanics in
the early Universe. The approach is associated with the density matrix
deformation and not with the deformation of Heisenberg algebra \textbf{Blum
(1981).} The density matrix $\rho $ is the statistical operator that may be
used to calculate the average for any physical quantity in quantum
mechanics. This operator was introduced by J. Neuman and L.D. Landau in
1927.

Let us consider inequalities (\ref{eq8}) and (\ref{eq9}) more closely. The distinguishing
feature of the first inequality is its linearity, i.e. all the involved
quantities are of the first order, whereas the second inequality is
quadratic involving the squares. From the viewpoint of mathematics, in the
case under consideration there exists a ``minimum length''
\begin{equation}
\label{eq12}
\Delta x\ge \ell _{\min } \sim \ell _p ,
\end{equation}
and every coordinate measurement may be performed to an accuracy that is not
in excess of a particular minimum length $\ell _{\min } $ of the order of
Planck's length. Also, this suggests that there is a maximum energy on the
order of Planck's energy$E_{\max } \sim E_p $.

Thus, a distinctive feature of quantum mechanics of the Early Universe is
the existence of a minimum length,$\ell _{\min } \sim \ell _p $, that is
referred to as the fundamental length too.

For our further consideration the existence of a minimum length is the
starting point.

Since in quantum mechanics the measuring procedure is determined by the
density matrix, the problem is as follows. Provided quantum mechanics
involves the fundamental length, the question is: ``How does the density
matrix on the retention of the measuring procedure change?'' In this case
the measuring procedure represents an algorithm used to calculate the
averages of operators. As demonstrated in previous works
\textbf{(Shalyt-Margolin and Suarez 2003, Chap. 3), (Shalyt-Margolin 2005,
Chap. 3), }one of the approaches to this problem is associated with
the density matrix deformation $\rho $. In so doing $\rho $ becomes
dependent on a new small dimensionless parameter$\alpha =\ell _{\min }^2
/\ell ^2$, where $\ell _{\min } $- minimum length, and $\ell $ - measuring
scale determined by the energy measured. The parameter $\alpha $ may be
represented in terms of energy $\alpha =E^2/E_{\max }^2 $, where $E$ -
measured energy, $E_{\max } \sim E_p $ - energy of the process. This new
parameter is measured over the interval$0<\alpha \le 1/4$. At low energies
$\rho \left( \alpha \right)$ becomes the density matrix of the well-known
quantum mechanics $\rho $
\begin{equation}
\label{eq13}
\mathop {\lim }\limits_{\alpha \to 0} \rho \left( \alpha \right)=\rho .
\end{equation}
Note that the parameter $\alpha $ includes all the above-mentioned
fundamental constants ($\hbar $, $c$, and $G)$ because $\ell _{\min } $ and
$E_{\max } $are expressed in terms of these constants.

Heisenberg algebra is subjected to deformation too. As this takes place in
the simplest and minimal variant, the first and second relations in (\ref{eq10})
remain invariable, whereas the last one is changed as follows:
\begin{equation}
\label{eq14}
\left[ {q_j ,p_k } \right]=i\hbar \lambda \left( \alpha \right)\delta _{jk}
,
\end{equation}
where $\lambda \left( \alpha \right)$ - function of $\alpha $ characterized
by the property
\begin{equation}
\label{eq15}
\mathop {\lim }\limits_{\alpha \to 0} \lambda \left( \alpha \right)=1.
\end{equation}
The function $\lambda \left( \alpha \right)=\exp (-\alpha )$, in particular,
meets this property.

So, we actually get deformation of Heisenberg algebra that changes to the
conventional Heisenberg algebra at low energies.

But to derive the generalized uncertainty relations (\ref{eq9}), the foregoing
``minimal'' variant of the Heisenberg algebra deformation is insufficient
- an extended variant is required, where $\left[ {q_i ,q_j } \right]\ne 0$
as in formula (\ref{eq11}). What form takes this ``non-minimal'' deformation variant
of Heisenberg algebra is presently unknown and remains to be found.

\section{Deformed statistical theory of the Early Universe }
As noted in the preceding section, the uncertainty relations may be modified
(deformed) in the quantum mechanics of the early Universe.

Recall that the pair $\left( {p,x} \right)$ in formulae (\ref{eq8}), (\ref{eq9}) is called
the conjugate pair of variables in quantum mechanics \textbf{Messiah (1965,
Chap. 4, Sect 2)}. Besides, there is another conjugate pair $\left( {E,t}
\right)$ and hence in terms of this pair (\ref{eq8}) has an analog \textbf{Messiah
(1965, Chap. 4, Sect }2)
\begin{equation}
\label{eq16}
\Delta E\Delta t\ge \hbar .
\end{equation}
Relation (\ref{eq16}) relates the uncertainty $\Delta E$ of the value assumed by
this dynamic variable to the time interval $\Delta t$ characteristic for the
time evolution of a system.

Relation (\ref{eq16}) possesses an explicit physical meaning: the energy measurement
accuracy $\Delta E$ is related to the time $\Delta t$ required for this
measurement. In particular, (\ref{eq16}) indicates that the lower the time interval
the poorer the measuring accuracy. However, relations (\ref{eq16}) and (\ref{eq8}) are
distinguished by one more fundamental feature. In (\ref{eq8}) both conjugate
variables $p$ and $x$ are operators of quantum mechanics, whereas in (\ref{eq16})
only energy $E$ is such an operator. Both (\ref{eq16}) and (\ref{eq8}) are modified at
Planck's scales (in the early Universe) \textbf{Shalyt-Margolin and
Tregubovich (2004, p.73), Shalyt-Margolin (2005, p.62)}
\begin{equation}
\label{eq17}
\Delta t\ge \frac{\hbar }{\Delta E}+\gamma t_p^2 \frac{\Delta E}{\hbar }.
\end{equation}
And both modified relations (\ref{eq9}) and (\ref{eq17}) may be written in the canonical
form as
\begin{equation}
\label{eq18}
\left\{ {\begin{array}{l}
 \Delta x\ge \frac{\hbar }{\Delta p}+\gamma \left( {\frac{\Delta p}{P_{pl}
}} \right)\frac{\hbar }{P_{pl} } \\
 \Delta t\ge \frac{\hbar }{\Delta E}+\gamma \left( {\frac{\Delta E}{E_p }}
\right)\frac{\hbar }{E_p } \\
 \end{array}} \right.\quad ,
\end{equation}
where $P_{pl} =E_p /c=\sqrt {\hbar c^3/G} $ - Planck's momentum.

Now we consider the thermodynamic uncertainty relation between the inverse
temperature and interior energy of a macroscopic ensemble\textbf{ Lavenda
(1991, Chap.4, Sect.4.9)}
\begin{equation}
\label{eq19}
\Delta \frac{1}{T}\ge \frac{k}{\Delta U}
\end{equation}
where $k$ is the Boltzmann constant.

\textbf{Bohr (1932)} and \textbf{Heisenberg (1969)} were the first to point
out that such kind of uncertainty principle should take place in
thermodynamics.

The uncertainty in thermodynamic measurements occurs as follows
\textbf{Lavenda (1991, Chap.4, Sect.4.9)}: ``Let us think of the simplest
case when a system is brought in contact with a thermal reservoir. Energy
ceases to be a thermodynamic function, becoming a random variable instead,
as the value of all the external parameters determining this system is
insufficient to find the energy. And energy of the system fluctuates.
Observations of the energy may be used for the estimation of its conjugate
intensive quantity. As the thermostat size is decreased, the energy
measurements become significantly more accurate, and a certain energy value
appears in the limit. On the other hand, measurements of energy become less
accurate with a growing thermostat size. Therefore, a particular temperature
of a system may be determined in the limit of an infinite thermostat. An
infinite thermal reservoir means an infinite heat capacity. As the heat
capacity is inversely proportional to the dispersion of thermal
fluctuations, in the limit it goes to zero.''

Since the thermodynamic uncertainty relation (\ref{eq19}) have been proved by many
authors and in various ways \textbf{(Lavenda 1991, Chap.4; Uffink and van
Lith-van Dis 1999 and references in them)}, their validity is
unquestionable. Nevertheless, relation (\ref{eq19}) was established using a standard
model for the infinite-capacity heat bath encompassing the ensemble. But it
is obvious from the above inequalities that at very high energies the
capacity of the heat bath can no longer be assumed infinite at the Planck
scale. Indeed, the total energy of the pair heat bath - ensemble may be
arbitrary large but finite, merely as the Universe is born at a finite
energy. Thus, the quantity that can be interpreted as a temperature of the
ensemble must have the upper limit and so does its main quadratic deviation.
In other words, the quantity $\Delta (1/T)$ must be bounded from below. But
in this case an additional term should be introduced into (\ref{eq19})
\textbf{Shalyt-Margolin and Tregubovich (2004, p.74), Shalyt-Margolin (2005,
p.68)}
\begin{equation}
\label{eq20}
\Delta \frac{1}{T}\ge \frac{k}{\Delta U}+\eta \,\Delta U
\end{equation}
where $\eta $ is a coefficient. Dimension and symmetry reasons give
\[
\eta \sim \frac{k}{E_p^2 }.
\]
Similar to the previous cases, inequality (\ref{eq20}) leads to the fundamental
(inverse) temperature.
\begin{equation}
\label{eq21}
T_{max} =\frac{\hbar }{\Delta t_{min} k},\quad \beta _{min}
=\frac{1}{kT_{max} }=\frac{\Delta t_{min} }{\hbar }.
\end{equation}
Thus, we obtain the following system of generalized uncertainty relations in
the symmetric form
\begin{equation}
\label{eq22}
\left\{ {{\begin{array}{*{20}c}
 {\Delta x} \hfill & \ge \hfill & {\frac{\hbar }{\Delta p}+\gamma \left(
{\frac{\Delta p}{P_{pl} }} \right)\,\frac{\hbar }{P_{pl} }+...} \hfill \\
 \hfill & \hfill & \hfill \\
 {\Delta t} \hfill & \ge \hfill & {\frac{\hbar }{\Delta E}+\gamma \left(
{\frac{\Delta E}{E_p }} \right)\,\frac{\hbar }{E_p }+...} \hfill \\
 \hfill & \hfill & \hfill \\
 {\Delta \frac{1}{T}} \hfill & \ge \hfill & {\frac{k}{\Delta U}+\gamma
\left( {\frac{\Delta U}{E_p }} \right)\,\frac{k}{E_p }+...} \hfill \\
\end{array} }} \right.
\end{equation}
or in the equivalent form
\begin{equation}
\label{eq23}
\left\{ {{\begin{array}{*{20}c}
 {\Delta x} \hfill & \ge \hfill & {\frac{\hbar }{\Delta p}+\gamma l_p
^2\frac{\Delta p}{\hbar }+...} \hfill \\
 \hfill & \hfill & \hfill \\
 {\Delta t} \hfill & \ge \hfill & {\frac{\hbar }{\Delta E}+\gamma t_p
^2\frac{\Delta E}{\hbar }+...} \hfill \\
 \hfill & \hfill & \hfill \\
 {\Delta \frac{1}{T}} \hfill & \ge \hfill & {\frac{k}{\Delta U}+\gamma
\frac{1}{T_p ^2}\frac{\Delta U}{k}+...} \hfill \\
\end{array} }} \right.
\end{equation}
Here $T_p $ is the Planck temperature$T_p =E_P /k$, dots in the right side
of (\ref{eq22}) and (\ref{eq23}) denote terms of the higher-order smallness. Also, it is
assumed that the factor $\eta $ in the right side of (\ref{eq20}) is equal to
$\gamma k/E_p^2 =\gamma /kT_p^2 $. Note that the last-mentioned inequality
is symmetrical to the second one with respect to substitution
\[
t\mapsto \frac{1}{T},\hbar \mapsto k,\Delta E\mapsto \Delta U.
\]
However, this observation can by no means be regarded as a rigorous proof of
the generalized uncertainty relation in thermodynamics.

There is reason to believe that rigorous justification for the latter
(thermodynamic) inequalities in systems (\ref{eq22}) and (\ref{eq23}) may be made by means
of a certain deformation of Gibbs distribution.

The generalized uncertainty relations in thermodynamics (\ref{eq22}) substantiated
by the author \textbf{Shalyt-Margolin and Tregubovich (2004, p.74),
Shalyt-Margolin (2005, p.68) }still require a strict mathematical proof, but
by the present time the result has been citated in the monograph
\textbf{Carroll (2006)}.

In the previous work \textbf{Shalyt-Margolin }and\textbf{ Tregubovich (2004,
pp.78--80), Shalyt-Margolin (2005, pp.62--67) }the density matrix
deformation method has been developed as applied to statistical
mechanics at Planck's temperature. Specifically, it has been proved
that statistical mechanics of the early Universe is also modified (deformed)
as compared to the well-known statistical mechanics. In this modified
statistical mechanics, in particular, any ensemble has a temperature that
could not be higher than some maximum temperature $T_{\max } \sim T_p $.
Because of this, the statistical density matrix for very high temperatures
is deformed too. It begins to be dependent on the new dimensionless
parameters $\tau =T^2/T_{\max }^2 $, where $T$ - ensemble temperature
\[
\rho _{stat} \to \rho _{stat} \left( \tau \right)
\]
\[
\mathop {\lim }\limits_{\tau \to 0} \rho _{stat} (\tau )=\rho _{stat} .
\]
Similar to the parameter $\alpha $, the parameter $\tau $ is varying over
the interval $0<\tau \le 1/4$. At low temperatures, passage to the limit is
also valid with the density matrix $\rho _{stat} $ for the canonical Gibbs
distribution in right hand side of the last formula.

Thus, compared to the well-known statistical mechanics, statistical
mechanics of the early Universe (i.e. at super high temperatures of the
order of the Planck's temperature) involves a new small parameter, varying
over the same interval as the corresponding new small parameter introduced
in quantum mechanics at Planck's scales. In this case the statistical
density matrix is also deformed so that, in the limit of low temperatures,
the statistical density matrix appears corresponding to the canonical Gibbs
distribution.

\section{Entropy in the present and early Universe }
Because quantum and statistical mechanics of the early Universe are modified
(deformed) compared to the conventional ones at the known energies, the
notions involved are also modified, specifically, being dependent on new
parameters.

This is true for the entropy notion as well. In this case we use the
standard understanding of entropy from the information theory (e.g., see
\textbf{Adami (2004) pp. 1-3}):

``The concepts of entropy and information quantify the ability of
observers to make \textit{predictions}, in particular how well an
observer equipped with a specific measurement apparatus can make
predictions about another physical system. Shannon entropy (also
known as \textit{uncertainty}) is defined for mathematical objects
called \textit{random variables}. A random discrete variable $X$
is a variable that can take on a finite number of discrete states
$x_i $, where $i=1,...,N$ with probabilities $p_i $. Now, physical
systems are not mathematical objects, nor are their states
necessarily discrete. However, if we want to quantify our
uncertainty about the state of a physical system, then in reality
we need to quantify our uncertainty about the \textit{possible
outcomes of a measurement of that system}. Our maximal uncertainty
about a system is not a property of the system, but rather a joint
property of the measurement device and the device with which we
are about to examine the system.\textbf{ }If our measurement
device, for example, is simply a ``presence-detector'' then the
maximal uncertainty we have about the physical system under
consideration is 1 bit, which is the amount of \textit{potential
information} we can obtain about that system. Thus, the entropy of
a physical system is undefined if we do not specify the device
that we are going to use to reduce that entropy. Here we consider
only the \textit{discrete} version of the Shannon entropy, which
is given in terms of the probabilities $p_i $ as \textbf{(Shannon
(1948))}:
\begin{equation}
\label{eq24} H(X)=-\sum_{i=1}^{N} p_{i}\log p_{i}.
\end{equation}
For any physical system, how are those probabilities obtained? In principle,
this can be done both by experiment and by theory. Once we have defined the
$N$ possible states of the system by choosing a detector for it, the \textit{a priori}
maximal entropy, corresponding to the uniform distribution (all states
equally likely) is then
\begin{equation}
\label{eq25}
H_{\mbox{max}} =\log N.
\end{equation}
Classical experiments using detector can now sharpen our knowledge
of the system.\textbf{~}By tabulating the frequency with which
each of the $N$ states appears, we can estimate the probabilities
$p_i $. Note, however, that this leads to a biased estimate of the
entropy (\ref{eq24}), that approaches its~true value~only in the
limit of an infinite number of trials. On the other hand, some of
the possible states of the system (or more precisely, possible
states of the detector interacting with the system) can be
eliminated by using some knowledge of the physics of the system.
For example, we may have some initial data about the system. This
becomes clear in particular if the degrees of freedom that we
choose to characterize the system with are position, momentum, and
energy, i.e., if we consider the \textit{thermodynamical entropy}
of the system'' \textbf{ Feynman (1972, Chap. 1)}.

The notion of entropy for the quantum system $A$ was generalized by von
Neumann in 1932 as follows\textbf{:}
\begin{equation}
\label{eq26}
S\left( {\rho _A } \right)=-Tr\rho _A \log \rho _A ,
\end{equation}
where $\rho _A $ - density matrix of the system $A$, and $\log \rho _A $ is
understood as such an operator that $e^{\log \rho _A }=\rho _A $.

But as indicated in the preceding section, new parameters are introduced
into a quantum theory of the early Universe. Specifically, in Section 3 it
is demonstrated that the deformation of the density matrix is associated
with the occurrence of the new parameter$\alpha $.

Note that formula (25) may be extended as $\alpha $ may be included twice
\textbf{Shalyt-Margolin (2004, 1, p.397, 2004, 2, p.2040)}
\begin{equation}
\label{eq27}
S_{\alpha _2 }^{\alpha _1 } =-Tr\rho \left( {\alpha _1 } \right)\log \rho
\left( {\alpha _2 } \right)
\end{equation}
where$0<\alpha _1 ,\alpha _2 \le 1/4$.

Physically, $S_{\alpha _2 }^{\alpha _1 } $ may be interpreted as follows.
This is a two-dimensional entropy density calculated at the scales (or
energies) associated with the deformation parameter $\alpha _2 $ by the
observer who is at the energies specific for the deformation parameter
$\alpha _1 $.

Contrary to the classical quantum mechanics, this quantity is no longer
scalar, but rather a matrix value. The associated matrix seems to be
asymmetric
\begin{equation}
\label{eq28}
S_{\alpha _2 }^{\alpha _1 } \ne S_{\alpha _1 }^{\alpha _2 } \quad .
\end{equation}
The conventional notion of statistical entropy appears in the limit $\alpha
_1 \to 0$, $\alpha _2 \to 0$
\begin{equation}
\label{eq29}
S_0^0 =S=-Tr\rho \log \rho \quad .
\end{equation}
In this manner the notion of entropy in the early Universe is more splendid
and complicated, including a matrix instead of the number.

In the author's works \textbf{Shalyt-Margolin (2004, 1), Shalyt-Margolin
(2004, 2)} it has been indicated that the matrix $S_{\alpha _2 }^{\alpha _1
} $ may be used for the solution of several problems in a theory of black
holes. With this matrix, in particular, one can obtain
\textbf{Shalyt-Margolin (2004, 1, p.397), Shalyt-Margolin (2004, 2, p.2043)}
the well-known Bekenstein -- Hawking formula \textbf{Bekenstein (1973)} for
the entropy of a black hole in the semi-classical approximation
\begin{equation}
\label{eq30}
S_{BH} =\frac{A}{4\ell _p^2 },
\end{equation}
where $A$ - surface area of the event horizon of a black hole. In
the works mentioned the author has developed an approach to solve
the Hawking's information paradox problem \textbf{Hawking (1976)}
for black holes using $S_{\alpha _2 }^{\alpha _1 } $.

As demonstrated, when this problem is considered in terms of the introduced
matrix (entropy density matrix), there is no information loss at the black
hole as, with respect to the infinitely remote observer, the information
concerning the initial singularity and black hole singularity is the same
\textbf{Shalyt-Margolin (2004, 1, Chap. 4), Shalyt-Margolin (2005, pp.72
--74)}
\begin{equation}
\label{eq31}
 S_{1/4}^0 =S_{1/4}^0 \quad .
\end{equation}
Besides, it is demonstrated that a series expansion of the matrix element
$S_{1/4}^\alpha $in terms of small parameter $\alpha $ may give quantum
corrections for the semi-classical value of the black hole entropy in the
right side of formula (29) \textbf{Shalyt-Margolin (2006)}.

Nevertheless, note that in the above-mentioned work \textbf{Shalyt-Margolin
(2006)} the calculation of quantum corrections with the use of the deformed
density matrix has been contemplated rather than developed. By the present
time, the approach intended to study thermodynamics of black holes, and
entropy in particular, using the GUP has been better developed. In an
earlier work \textbf{Medved, Vagenas, (2004) }GUP has been used to obtain an
exact value of the logarithmic correction for entropy of a black hole ; and
also the calculation of higher-order corrections has been planned. Based on
GUP, in a later work \textbf{Bolen, Cavaglia, (2005) }the black hole
thermodynamics has been studied in de Sitter and Anti-de Sitter spaces. A
group of authors \textbf{Cardoso, Berti, Cavaglia, (2005) }has reviewed
different methods to estimate the total gravitational energy emitted in
 higher-dimensional scenarios allowing for the formation of
mini-black holes from TeV-scale particle collisions. Of course, GUP should
play an important part in such processes. Finally, it has been shown
\textbf{Adler, Chen, Santiago, (2001), Chen, Adler (2003)} and \textbf{Chen
(2003)} that, owing to GUP, a black hole is evaporated incompletely, having
the stable remainder with a mass on the order of the Planck mass. In this
work it is suggested that the remainders might form the basis for Dark
Matter; considering the validity of GUP, the problems of correcting the
black hole temperature are treated.

The notion of symmetry in the deformed theories due to the introduction of
new parameters becomes wider. In his work the author \textbf{Shalyt-Margolin
(2004, 1, Chap. 3)} is concerned with the unitary symmetry problem in the
early Universe using the parameter$\alpha $. This symmetry group on passage
to the limit for $\alpha \to 0$ should produce the infinite unitary group
$U$of quantum mechanics and quantum field theory. The quantum field theory
is mentioned here, as in some papers \textbf{Shalyt-Margolin (2004, 3),
Shalyt-Margolin (2005, 2) }the author has demonstrated\textbf{
}that from the deformed quantum mechanics one can proceed to the deformed
quantum field theory. Certainly, at such symmetries the density matrix of
entropy $S_{\alpha _2 }^{\alpha _1 } $ should be retained.

Note that the approach to a quantum theory of the early Universe with the
use of the Heisenberg's algebra deformation necessitates extension of the
symmetry notion. Specifically, this approach involves the quantum groups
\textbf{Maggiore (1994)} which also represent the deformed algebraic
objects, not the groups in the sense of the standard definition \textbf{Weyl
(1931)}.

\section{New concepts in fundamental physics. Holographic principle. }
Perfectly new concepts have appeared in fundamental physics in the last
fifteen years as regards the quantity of information and entropy contained
in cosmological objects and in the Universe as a whole. The quintessence of
these concepts is the Hooft-Susskind Holographic Principle that may be
formulated in its simplest form as follows \textbf{Bousso (2002, Chap.3,
Sec. C):}

``\textbf{The region }$V$\textbf{ with boundary of }$A$\textbf{ is fully
described by no more than }$A/4\ell p^2$\textbf{ degrees of freedom, or
about 1 bit of information per Plank area. A fundamental theory, unlike
local field theory, should incorporate this counterintuitive result''.}

This means that entropy of the region $V$ satisfies the inequality
\begin{equation}
\label{eq32}
S\left( V \right)\le \frac{A}{4\ell _p^2 }.
\end{equation}
The Holographic principle provides an answer for the question: ``How many
degrees of freedom are there in nature, at the most fundamental level?'' The
Principle has been first put forward in the works of \textbf{'t
Hooft}\textbf{ (1993), (2000) }and\textbf{ Susskind (1995).}

It should be noted that, as the foregoing formulation presents the principle
and not a physical law, we can consider only the objects meeting this
principle and the conditions for which this principle is valid.

Initially, the Holographic Principle has been substantiated \textbf{Bousso
(2002, Chap.2, Sec. C1, Susskind process): } ``Let us consider an isolated
matter system of mass $E$ and entropy $S$ residing in the space-time ${\rm
M}$. It is assumed that the asymptotic structure of ${\rm M}$ permits the
formation of a black hole. For example, ${\rm M}$ is asymptotically flat.
And let $A$ be the area of the circumscribing sphere, i.e., the smallest
sphere that fits around the system. However $A$ is well-defined only if the
metric near the system is at least approximately spherically symmetric. This
will be the case for all spherically symmetric systems, and for all weakly
gravitating systems, but not for strongly gravitating systems lacking
spherical symmetry. Besides, we assume that the matter system is stable on a
timescale much greater than$A^{1/2}$. It persists and does not expand of
collapse rapidly, so that the time-dependence of $A$ will be negligible.''

Then, in accordance with the generalized second law, for black holes we have

\textbf{Bekenstein (1974)}
\begin{equation}
\label{eq33}
S_{matter} \le S_{BH} =\frac{A}{4\ell _p^2 },
\end{equation}
where $S_{BH} $ - entropy of a black hole with the submerged isolated matter
system whose surface area of the event horizon equals $A$. As has been
already mentioned, $A$ is well-defined only if the metric near the system is
at least approximately spherically symmetric.

In the nineties of the last century the Holographic Principle has been
proved for a considerably wider class of geometries \textbf{Bousso (2002,
Chap. 5, 6).}

The Holographic Principle necessitates radical reconsideration of the
existing viewpoints in fundamental physics regarding the degrees of freedom
of some system, because the number of degrees of freedom $N$calculated from
the local quantum field theory for the system within the volume $V$ is
proportional to $V$ \textbf{Bousso (2002, Chap. 3, Sec. C)}
\begin{equation}
\label{eq34}
N\sim V.
\end{equation}
Actually, the Holographic Principle is an important step in the
direction of a unified theory embracing gravity and quantum field
theory. Moreover, the Principle leads to the effective
dimensionality reduction due to the fact that by this principle
all information about the object is concentrated on the surface
and hence its effective dimensionality is less by one.
Specifically, the 3D stationary objects meeting the Holographic
Principle are virtually specified by their 2D boundary. In the
general theory similar situation is observed with a random number
of measurements: a physical system given in the volume $V$ at the
$n$-metric manifold ${\rm M}$ and meeting the Holographic
Principle is determined by its $(n-1)$-metric boundary $A$.
Naturally, a symmetry of $\tilde {G}\left( V \right)\to G\left( A
\right)$, where $\tilde {G}\left( V \right)$ is the initial
symmetry group of the system within the volume $V$, and $G\left( A
\right)$ - corresponding symmetry group of the same system
projected to the boundary $A$.

Finally, it should be noted that the initially formulated Holographic
Principle has been recently generalized \textbf{Bousso (2002, Chap. 8, Sec.
B)} to assume that \textbf{`` {\ldots} it is a law of physics which must be
manifested in the underlying theory. This theory must be a unified quantum
theory of matter and space-time''.}

\section{Conclusion}
In conclusion, we revert to the stable deformation that has been
considered at the beginning. The Heisenberg's algebra deformations
are introduced due to the involvement of GUP and minimal length in
quantum mechanics. These deformations are stable in the sense of
\textbf{Definition 2 }given in\textbf{ Section 1.} But this is not
true for the unified algebra of Heisenberg and Poincar\'{e}. This
algebra does not carry the indicated immunity. It is suggested
that the Lie algebra for the interface of the gravitational and
quantum realms is in its stabilized form. Now it is clear that
such a stability should be raised to the status of a physical
principle. In a very interesting work of \textbf{Ahluwalia --
Khalilova (2005) }it has been demonstrated that the stabilized
form of the Poincar\'{e}-Heisenberg algebra  \textbf{Vilela-Mendes
(1994), Chryssomalakos, Okon (2004)} carries three additional
parameters: ``a length scale pertaining to the Planck/unification
scale, a second length scale associated with cosmos, and a new
dimensionless constant with the immediate implication that `point
particle' ceases to be a viable physical notion. It must be
replaced by objects which carry a well-defined, representation
space dependent, minimal spatiotemporal extent''.


\end{document}